\documentclass[twocolumn,aps,pra,showpacs,amsmath,amssymb,superscriptaddress,showkeys,longbibliography]{revtex4-1}

\usepackage{appendix}
\usepackage{xcolor}
\usepackage{hyperref}
\usepackage{graphicx}

\begin{document}
\title{An Effective-Current Approach for Hall\'{e}n's Equation in Center-Fed Dipole Antennas with Finite Conductivity}
\author{Themistoklis K. Mavrogordatos}
\email[email: ]{themis.mavrogordatos@fysik.su.se}
\affiliation{Department of Physics, AlbaNova University Center, Stockholm University, SE-106 91, Stockholm, Sweden}
\author{Anastasios Papathanasopoulos}
\email[email: ]{anastasispapath@g.ucla.edu}
\affiliation{Department of Electrical Engineering, University of California, Los Angeles, Los Angeles, CA 90095, USA}
\author{George Fikioris}
\email[email: ]{gfiki@ece.ntua.gr}
\affiliation{Department of Electrical and Computer Engineering, National Technical University of Athens, Athens, 15780 Zografou, Greece}

\date{\today}

\begin{abstract}
We propose a remedy for the unphysical oscillations arising in the current distribution of carbon nanotube and imperfectly conducting antennas center-driven by a delta-function generator when the approximate kernel is used. We do so by formulating an effective current, which was studied in detail in a 2011 and a 2013 paper for a perfectly conducting linear cylindrical antenna of infinite length, with application to the finite-length antenna. We discuss our results in connection with the perfectly conducting antenna, providing perturbative corrections to the current distribution for a large conductance, as well as presenting a delta-sequence and the field of a Hertzian dipole for the effective current in the limit of vanishing conductance. To that end, we employ both analytical tools and numerical methods to compare with experimental results.
\end{abstract}


\maketitle

\section{Introduction}

{\bf T}he most fundamental quantity to be determined theoretically for a linear antenna is arguably the current distribution $I(z)$ along its axis. The function $I(z)$ satisfies an integrodifferential equation of the Pocklington type (eqn. \ref{eq:Pocklington} below) or the equivalent integral equation (eqn. \ref{eq:Hallen} below) commonly referred to as Hall\'{e}n's equation. These two equations assume an antenna which is center-driven by the ``delta-function generator (DFG)'', which is the simplest yet still widely employed type of feed. Once $I(z)$ is known, the rest of the essential electrical characteristics of the linear antenna, such as input impedance and radiation efficiency, can be readily derived and subsequently compared with experimental data.

Hall\'{e}n's integral equation is usually solved by applying the method of moments (MoM) \cite{harrington1993field}. This equation assumes two forms depending on the choice of the kernel, namely the ``exact'' and the ``approximate'' (or ``reduced'') kernel. Although the exact kernel can be efficiently employed \cite{bruno2007regularity}, the much simpler approximate kernel still features prominently \cite{FundamentalTransm, keller2014electromagnetic, huang2008performance, fichtner2008investigation, attiya2009lower, forati2011new}. For the perfectly conducting (PC) antenna, the use of the approximate kernel is extensively discussed in the modern literature \cite{kraus, balanis_em, stutzman, balanis_antenna}. The use of the approximate kernel in conjunction with small discretization length, i.e. for a sufficiently large number of basis functions, necessarily entails the appearance of unphysical oscillations as an important consequence of the non-solvability of Hall\'{e}n's equation \cite{Hallen2001, AntennaTheoryCh8, FGG12013}. 

A remedy for the pathologies attendant to the numerical solution of Hall\'{e}n's equation, when formulated in terms of the widely-used approximate kernel, has been recently introduced. In particular, an effective-current method has been proposed to post-process the oscillatory current distribution obtained with the approximate kernel \cite{PapakanellosMAS,papakanellos2007possible}. The convergence of the method has been investigated in \cite{EffCurrent2011, EffCurrent2013}. The main results originate from an analytical study of the infinite antenna, which serves as a guide when investigating the behavior of the numerical solutions for the current distribution on the finite-length antenna.

A very recent paper \cite{anastasiscnt} extended the scope of the previous studies by considering finite rather than infinite conductance, motivated by current work on optical antennas \cite{alu2013theory, bharadwaj2009optical, alu2008input} and particularly carbon nanotubes (CNTs) \cite{FundamentalTransm, wrongconclusion, hanson2006current, xiapeiro, keller2014electromagnetic, huang2008performance, fichtner2008investigation}. These antennas are modeled as imperfect conductors, where finite conductance is an essential consideration \cite{FundamentalTransm}. When the widely-employed approximate kernel is considered, unphysical oscillations may appear as well appear as well for center-fed antennas with finite conductance, as the authors of \cite{anastasiscnt} demonstrated. Their analysis extended to the infinite antenna in order to derive analytical asymptotic relations. While there are many similarities with the PC case, new difficulties are also encountered when the resistivity is appreciable. In the present work, we build upon these results, developing the effective-current method to eliminate the occurring oscillations and connect to the current distribution obtained for the exact kernel. Our main conclusions are also drawn by means of an asymptotic study for the infinite antenna. We provide connections to the actual finite-length antenna, and account for configurations of experimental interest.

\section{Integral equations and kernels}

In this introductory section, we will define the well-known central equations of our analysis yielding the current distribution on a linear cylindrical antenna driven by a DFG. We assume a $e^{-i\omega t}$ time dependence throughout. Our notation and conventions are very close to those of \cite{EffCurrent2011}, \cite{EffCurrent2013} and \cite{anastasiscnt}. For an antenna of radius $a$ and finite length $2h$, Pocklington's and Hall\'{e}n's equation are
\begin{equation}\label{eq:Pocklington}
\left(\frac{\partial^2}{\partial z^2} + k^2\right)\int_{-h}^{h} K(z-z^{\prime})I(z^{\prime})dz^{\prime}=\frac{iV k}{\zeta_0}\delta(z),
\end{equation}
and 
\begin{equation}\label{eq:Hallen}
\int_{-h}^{h} K(z-z^{\prime})I(z^{\prime})dz^{\prime}=\frac{iV}{2\zeta_0}\sin(k|z|)+C \cos(kz),
\end{equation}
respectively, where $-h<z<h$. In the above equations, $V$ is the voltage maintained at the driving point $z=0$, $k=2\pi/\lambda=\omega/c$ is the wavenumber, $\zeta_0=\sqrt{\mu_0/\epsilon_0}=376.73\,$Ohms is the free-space intrinsic impedance, and $\delta(z)$ is the Dirac delta function. In \ref{eq:Hallen}, the constant $C$ is determined by means of the boundary condition at the two ends of the antenna: 
\begin{equation}\label{eq:edgecondition}
I(\pm h)=0.
\end{equation}
The choice of kernel $K(z)$ in eqns. \ref{eq:Pocklington} and \ref{eq:Hallen} depends on the type of antenna under consideration. For the PC antenna, the exact kernel
\begin{equation}\label{eq:Exkernels}
K_{\rm ex}(z)=\frac{1}{8\pi^2} \int\limits_{-\pi}^{\pi} \frac{e^{ikR(z,\phi;a)}}{R(z,\phi;a)}d\phi,
\end{equation}
with $R(z,\phi;a)\equiv\sqrt{z^2+4a ^2\sin^2\left(\frac{\phi}{2}\right)}$, is often replaced by the so-called approximate kernel
\begin{equation}\label{eq:ApprKernel}
K_{\rm ap}(z)=\frac{1}{4\pi} \frac{e^{ik\sqrt{z^2+a^2}}}{\sqrt{z^2+a^2}},
\end{equation}
a simplification which leads to the appearance of unphysical oscillations of the current distribution at the ends of the antenna for the real part, and at both the center and the ends of the antenna for the imaginary part, as discussed in detail in \cite{Hallen2001}. A remedy of these oscillations for the center-driven antenna is discussed in \cite{EffCurrent2011, EffCurrent2013}. For a CNT and an imperfectly conducting antenna we also employ the kernel
\begin{equation}\label{eq:LossKernel}
K_{\text{loss}}(z)=\xi e^{ik|z|},
\end{equation}
such that the behavior of the antenna with practical interest is described by the kernel
\begin{equation}\label{eq:Kfull}
K_{\rm full, ap/ex}(z)=K_{\rm ap/ex}(z)+K_{\rm loss}(z). 
\end{equation}
In \ref{eq:LossKernel}, $\xi$ is a parameter with dimensions of inverse length (m$^{-1}$), related to the conductance $\sigma$ via the relation $\xi=1/(4\pi \zeta_0 a \sigma)$. For a PC antenna $\xi \to 0$, while in our analysis we will consider with special emphasis the limiting cases $\xi/k\gg 1$ and $\xi/k \ll 1$. Henceforth, we also let $\overline {F}(\zeta)=\int_{-\infty}^{\infty} F(z) e^{i\zeta z}\, dz$ stand for the Fourier transform (FT) of $F(z)$, which will be used in our analytical study.

\begin{figure*}[!t]
\centering
\includegraphics[width=0.9\paperwidth]{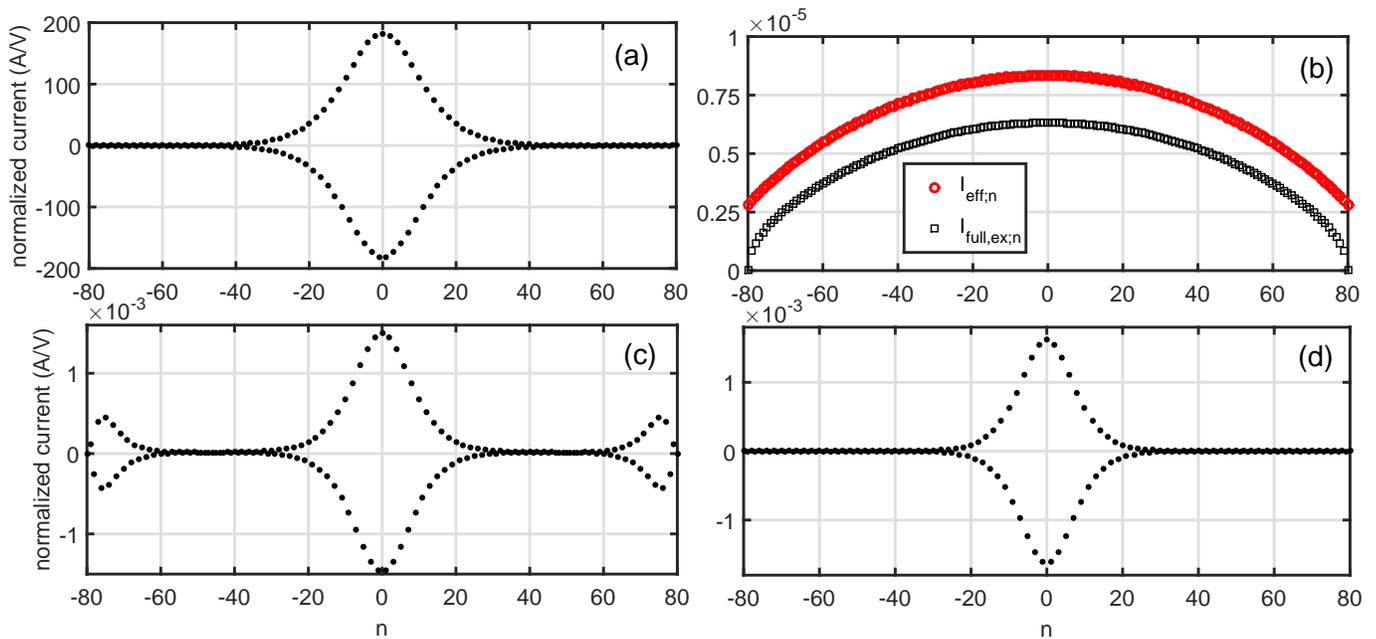}
\caption{Normalized current distributions (in $A/V$) for an imperfectly conducting antenna with $a/\lambda=0.005$, $\xi\lambda=7.1064\cdot 10^{-10}$ and $N=80$, after Table III of \cite{anastasiscnt}. In {\bf (a)} we plot $\mathrm{Re}\{I_{\rm full, ap;\,n}/V\}$ and in {\bf (b)} we plot $\mathrm{Re}\{I_{\rm eff;\,n}/V\}$ from \ref{eq:IeffInFin} at $\rho=a$ in red $\circ$, and $\mathrm{Re}\{I_{\rm full, ex;\,n}/V\}$ in black $\square$, for $h/\lambda=1/20$. In {\bf (c)} we plot $\mathrm{Re}\{I_{\rm full, ap;\,n}/V\}$ for $h/\lambda=1/15$. In {\bf (d)} we plot the effective current distribution from the analytical expression \ref{eq:FinalExpr} at $\rho=0$, with $N z_0=\lambda/15$. The distributions in frames (a, c) display the characteristic $(-1)^n$ oscillatory pattern, as captured by the asymptotic expression \ref{eq:FinalExpr} for the central oscillations due to finite conductivity.}
\label{fig:Figure1}
\end{figure*}

\section{Numerical methods for the finite-length antenna}\label{sec:Numerical}

In the present section, we will discuss the application of two different methods of moments (MoMs), employed to determine the current coefficients for an antenna with length $2h$ and radius $a$ such that $ka \ll 1$ and $a \ll h$ \cite{Hallen2001}. Central in all is the discretization in terms of $2N+1$ basis functions, together with a discretization length $z_0$ such that $N z_0 \simeq h$ for $N \gg 1$. 

\subsubsection{Method MoM-A} 
\begin{figure*}[!t]
\centering
\includegraphics[width=0.87\paperwidth]{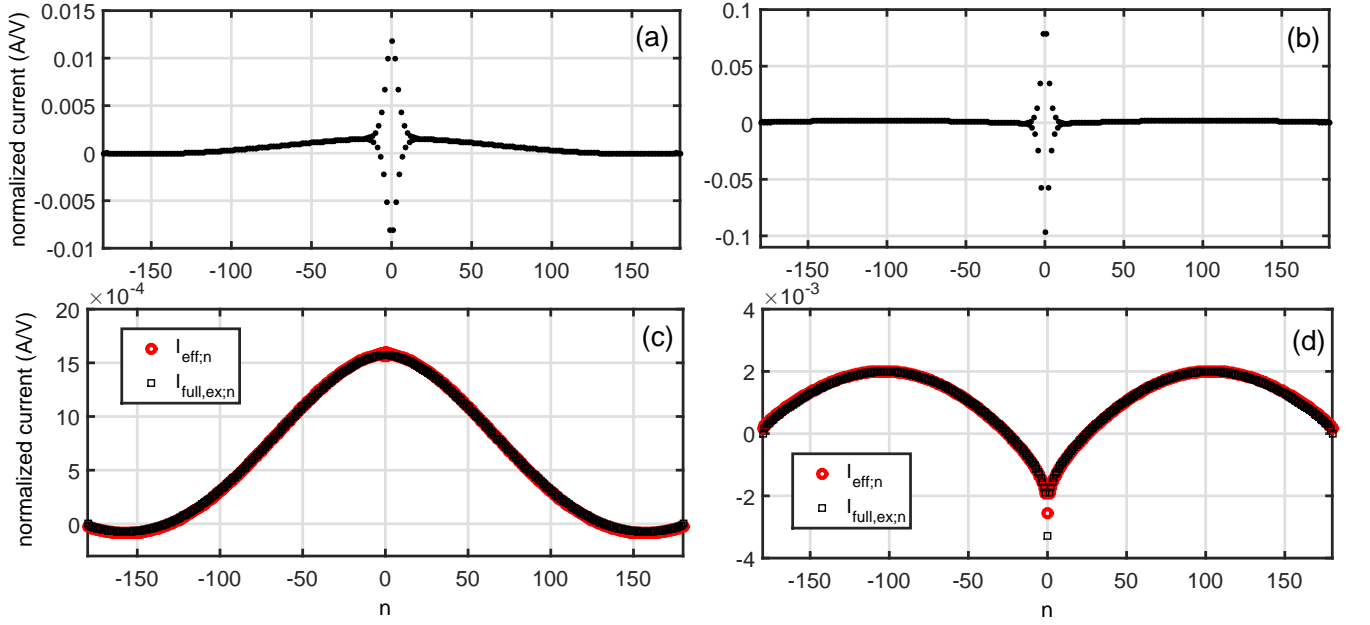}
\caption{Normalized current distributions (in $A/V$) for an imperfectly conducting antenna with $a/\lambda=0.007022$, $h/\lambda=1/2$, $\xi\lambda=0.4234$ and $N=180$, after Fig. 2 of \cite{PopovicTAP}. In {\bf (a), (b)} we plot $(\mathrm{Re, Im})\{I_{\rm full, ap;\,n}/V\}$ respectively. In {\bf (c), (d)} we plot $(\mathrm{Re, Im})\{I_{\rm eff;\,n}/V\}$, respectively, from \ref{eq:IeffInFin} at $\rho=a$ in red $\circ$, on top of $(\mathrm{Re, Im})\{I_{\rm full, ex;\,n}/V\}$, respectively, in black $\square$.}	
\label{fig:Figure2}
\end{figure*}

We seek the current distribution in the form 
\begin{equation}\label{eq:Itriang}
I(z) \simeq \sum_{n=-(N-1)}^{N-1} I_{n} t(z-nz_0),
\end{equation}
with $N z_0=h$. Here, $t(z)$ is the triangular basis function with length $2z_0$ such that $t(0)=1$ and $t(\pm z_0)=0$, defined as \cite{Hallen2001, EffCurrent2011}
\begin{equation}\label{eq:triangular}
t(z)=\begin{cases}\displaystyle\frac{z_0-|z|}{z_0},\quad -z_0\leq z \leq z_0, \\ 0, \quad |z|\geq z_0. \end{cases}
\end{equation}
Demanding that \ref{eq:Hallen} holds at the points $z=l z_0$, with $l=0, \pm 1,\ldots,\pm N-1$ produces the following system of equations for current coefficients $I_{n}$ \cite{Hallen2001}
\begin{equation}\label{eq:system1}
\sum_{n=-(N-1)}^{N-1} A_{l-n}I_{n}=\frac{iV z_0}{2\zeta_0}\sin(k|l|z_0)+z_{0} C\cos(klz_0),
\end{equation} 
in which the coefficients $A_{l}$ are given by the formula \cite{Hallen2001, anastasiscnt}
\begin{equation}\label{eq:Alcoeff}
A_{l}=A_{-l}=\int_{0}^{z_0}(z_0-z)[K(z+lz_0)+K(z-lz_0)]dz,
\end{equation}
in which $l=0, \pm 1, \pm 2,\ldots, \pm 2N$. In eqn. \ref{eq:Alcoeff} the choice of the kernel $K(z)$ depends on the particular configuration of interest according to eqns. \ref{eq:Exkernels}-\ref{eq:Kfull}. The constant $C$ is determined by \ref{eq:edgecondition}. The coefficients $I_{n}$ are then given by the superposition $I_{n}=I_{n}^{(1)}+C I_{n}^{(2)}$, where $I_{\pm N}=0$ and
\begin{equation}\label{eq:system2}
\sum_{n=-(N-1)}^{N-1}A_{l-n}I_{n}^{(1)}=B_{l}^{(1)}, \quad \sum_{n=-(N-1)}^{N-1}A_{l-n}I_{n}^{(2)}=B_{l}^{(2)},
\end{equation}
with $B_{l}^{(1)}=[iV z_0/(2\zeta_0)]\sin(k|l|z_0)$ and $B_{l}^{(2)}=z_{0} C\cos(klz_0)$. This method is referred to as Method B in \cite{Hallen2001}. Examples of the resulting oscillating  current distribution for $K=K_{\rm full}$ are depicted in Fig. \ref{fig:Figure1}(a, c) and Fig. \ref{fig:Figure2}(a, b). 

\subsubsection{Method MoM-B}

Alternatively, one can express the unknown current as a superposition of $2N+1$ step basis functions $u_{n}(z)$ with \cite{Hallen2001, anastasiscnt}
\begin{equation}\label{eq:pulsesbasis}
u_n(z) =
  \begin{cases}
    1&  \text{, if } \left( n-\frac{1}{2}\right)z_0<z<\left( n+\frac{1}{2}\right)z_0\\
    0&  \text{, otherwise,}\\
  \end{cases} 
\end{equation} 
where $(2N+1)z_0=2h$, substitute in \ref{eq:Hallen} and take the inner product with the same step functions $u_{n}$. This procedure results in the system of equations \ref{eq:system2} with new coefficients $B_{l}^{(1,2)}$ \cite{Hallen2001}. Results from the application of MoM-B have been obtained in \cite{anastasiscnt} for the imperfectly conducting finite-length antenna, in direct correspondence to Fig. \ref{fig:Figure1} here.  

\section{Effective current: definition and asymptotic analysis}\label{sec:effectcur}

The concept of an effective current draws inspiration from the Method of Auxiliary Sources (MAS). For the azimuthally-symmetric antenna, we can interpret the numerical solution as an auxiliary line current located on the $z$-axis, generating a magnetic field $H_{\phi}(\rho, z, z_0)$. The effective current at $\rho=a$ is then defined by the simple expression \cite{EffCurrent2013, FGG12013}
\begin{equation}\label{eq:defeffectA}
I_{\rm eff}(a,z,z_0)=2\pi a H_{\phi}(a, z, z_0)=-\left(\frac{2\pi \rho}{\mu_0} \right)\frac{\partial A_{z}}{\partial \rho}\Bigg|_{\rho=a},
\end{equation}
where $A_{z}$ is the $z$-component of the vector potential due to the line current. In this section, we present a numerical treatment for the infinite antenna, employing a different method from the ones discussed in Sec. \ref{sec:Numerical}. We denote by $I_{n}$, $n=0, \pm1, \pm 2, \ldots$, the basis functions coefficients, with $I_{0}$ located at $z=0$. We set the length $h$ in the limits of the integral in \ref{eq:Pocklington} to infinity ($h=\infty$), and we are looking for a current distribution of the form 
\begin{equation}\label{eq:CurrentSinus}
I(z) \simeq \sum_{n=-\infty}^{\infty} I_{n}^{(\infty)}s(z-nz_0),
\end{equation}
with the sinusoidal basis function $s(z)$ of length $2z_0$ defined as
\begin{equation}\label{eq:sinbasis}
s(z)=\begin{cases}\displaystyle\frac{\sin[k(z_0-|z|)]}{\sin(kz_0)},\quad -z_0\leq z \leq z_0, \\ 0, \quad |z|\geq z_0. \end{cases}
\end{equation}
Substituting \ref{eq:sinbasis} into \ref{eq:Pocklington} (with $h=\infty$) and taking the inner product with the (displaced) triangular functions $t(z-lz_0)$ yields an infinite system of equations for $I_{n}$. These coefficients are used in turn to define the \textit{effective current} via the magnetic field of the sinusoidal current $I_{n}^{(\infty)} s(z-nz_0)$ as \cite{EffCurrent2011, EffCurrent2013}
\begin{equation}\label{eq:IeffIn}
\begin{aligned}
I_{\rm eff}^{(\infty)}(\rho, z, z_0)&=\frac{1}{2i \sin(kz_0)}\\
&\times \sum_{n=-\infty}^{\infty}[f_{n+1}-2\cos(kz_0)f_{n}+f_{n-1}]\,I_{n}^{(\infty)},
\end{aligned}
\end{equation}
with $f_{n}\equiv\exp[ik\sqrt{(nz_0-z)^2+\rho^2}]$. Asymptotic expressions of $I_{\rm eff}(\rho,z,z_0)$ in the limit of small discretization length have been obtained in \cite{EffCurrent2011} to provide a link with the diverging exact-kernel current distribution for $\rho=a$ as $z\to 0$. 

We turn now to the case of the finite-length antenna for which the current coefficients are obtained numerically. Similarly to \ref{eq:IeffIn}, we write
\begin{equation}\label{eq:IeffInFin}
\begin{aligned}
&I_{\rm eff}(\rho, z=mz_0, z_0)=\frac{1}{2i \sin(kz_0)}\\
&\times \sum_{n=-(N-1)}^{N-1}[f_{n+1}-2\cos(kz_0) f_{n}+f_{n-1}]\,I_{n},
\end{aligned}
\end{equation}
where $I_{n}$ are the current coefficients resulting from the application of a MoM discussed in Sec. \ref{sec:Numerical}. Since $z_0$ is small, the extension of MoM-A to the infinite antenna (in which triangular instead of sinusoidal basis functions are used) yields virtually identical results (see Sec. 5 of \cite{EffCurrent2011}). The form \ref{eq:IeffInFin} has also been used to alleviate oscillations appearing in the solution of \ref{eq:Hallen} with a finite-gap generator \cite{FGG12013}. Following a series of manipulations, the effective current for the infinite antenna with a nonzero $z_0\ll a$ is found to be given by the expression
\begin{equation}\label{eq:Ieffdefinition}
I_{\rm eff}^{(\infty)}(\rho, nz_0,z_0)=\frac{1}{\pi}\int_{0}^{\pi}\frac{B(\theta, \rho, z_0)}{A(\theta, z_0)} d\theta,
\end{equation}
where the path of integration passes below $\theta=kz_0$ in the complex-$\theta$ plane. In \ref{eq:Ieffdefinition} we define [see. eqn. (2.1) of \cite{EffCurrent2013}]
\begin{equation}\label{eq:Bdef}
\begin{aligned}
B(\theta, \rho, z_0)&=B(-\theta, \rho, z_0)\equiv \frac{V z_0}{2\zeta_0}\sum_{l=-\infty}^{\infty}e^{ik\sqrt{(lz_0)^2+\rho^2}}e^{il\theta}\\
&=\frac{V}{2\zeta_0}\sum_{n=-\infty}^{\infty}\overline{L} \left(\frac{2\pi n-\theta}{z_0}, \rho\right),
\end{aligned}
\end{equation}
with
\begin{equation}\label{eq:LFourier}
\overline{L}(\zeta, \rho)=\frac{2k\rho}{i}\frac{1}{\sqrt{\zeta^2-k^2}}K_{1}\left(\rho \sqrt{\zeta^2-k^2}\right),
\end{equation}
in which $K_{1}$ is the usual modified Bessel function. We also define
\begin{equation}\label{eq:Atheta}
A(\theta, z_0)=\sum_{l=-\infty}^{\infty}A_{l}e^{il\theta},
\end{equation}
as the symbol of the infinite Toeplitz matrix $A_{ln}=A_{|l-n|}$, where the coefficients $A_{l}$ are given by \ref{eq:Alcoeff}. Their expression depends on the type of kernel under consideration.

\subsection{Limiting case of large conductance}\label{subsec:largecond}

We will now proceed to derive the correction to the asymptotic expressions of \cite{EffCurrent2013} due to finite conductance, reflected in the presence of the loss kernel, for sufficiently small $\xi/k$, as discussed in [Sec. VI. A of \cite{anastasiscnt}]. For $K=K_{\rm full, ap}=K_{\rm ap} + K_{\rm loss}$ \footnote{For simplicity we omit the subscript $({\rm full, ap})$ when referring to the effective current coefficients in the current section and in the figures where they feature.} we can make the approximation [eqn. (42) of \cite{anastasiscnt}]
\begin{equation}\label{eq:approxfirstordA}
\frac{1}{A_{\rm full}(\theta)}=\frac{1}{A_{\rm ap}(\theta) + A_{\rm loss}(\theta)}\simeq \frac{1}{A_{\rm ap}(\theta)}-\frac{A_{\rm loss}(\theta)}{[A_{\rm ap}(\theta)]^2}.
\end{equation}
The correction due to the second term in \ref{eq:approxfirstordA}, denoted by the subscript ${\rm corr}$ hereinafter, is given by \ref{eq:Ieff1} at the bottom of the page [see also eqn. (47) of \cite{anastasiscnt}].
\begin{widetext}
\begin{equation}\label{eq:Ieff1}
\begin{aligned}
&I_{\rm eff,\,corr}^{(\infty)}(\rho, nz_0,z_0)=-\frac{1}{\pi}\int_{0}^{\pi}\frac{B(\theta)A_{\rm loss}(\theta)}{[A_{\rm ap}(\theta)]^2}\, \cos(n\theta)\, d\theta\\
&\simeq \frac{(-1)^{n+1}}{16\pi z_{0}^2}\int_{0}^{\pi}\frac{A_{\rm loss}(\pi-\phi) B(\pi-\phi,\rho,z_0)[\cos^4(\phi/2)]^{-1}}{\left[(\pi-\phi)^{-2}\overline{K}_{\rm ap}\left(\frac{\pi-\phi}{z_0}\right) + (\pi+\phi)^{-2}\overline{K}_{\rm ap}\left(\frac{\pi+\phi}{z_0}\right)\right]^2}\, \cos(n\phi)\, d\phi.
\end{aligned}
\end{equation}
\end{widetext}
In the first integral of \ref{eq:Ieff1} we write $B(\theta)\equiv B(\theta, \rho, z_0)$, as defined in \ref{eq:Bdef}. Using the asymptotic relation $\overline{K}_{\rm ap}(\zeta)\sim e^{-a|\zeta|}/(2\sqrt{2\pi a |\zeta|})$ for $\zeta \to \pm \infty$ [see eqn. (1.11) of \cite{EffCurrent2013}] and effecting the change of variable $x=\phi a/z_0$ in the second integral of \ref{eq:Ieff1}, yields \ref{eq:Icorr2} also given at the end of the page.
\begin{widetext}
\begin{equation}\label{eq:Icorr2}
\begin{aligned}
&I_{\rm eff,\,corr}^{(\infty)}(\rho, nz_0,z_0)=\frac{(-1)^{n+1}}{2}\,e^{2\pi a/z_0}\\
&\times \int_{0}^{a/z_0}\frac{B\left(\pi - \frac{z_0 x}{a}\right)\left[kz_{0}^{2}\cos^4\left(\frac{z_0 x}{2a}\right)\right]^{-1} \left[k A_{\rm loss}\left(\pi - \frac{z_0 x}{a}\right)\right]}{\left[e^x\left(\pi - \frac{z_0 x}{a}\right)^{-5/2} + e^{-x}\left(\pi + \frac{z_0 x}{a}\right)^{-5/2} \right]^2}\, \cos\left(\frac{nz_0x}{a}\right)\, dx.
\end{aligned}
\end{equation} 
\end{widetext}

In the sole presence of the loss kernel we invoke eqn. (32) of \cite{anastasiscnt} to write
\begin{equation}\label{eq:Alossapp}
\begin{aligned}
&A_{\rm loss} \left(\pi - \frac{z_0 x}{a}\right)= -\frac{i\xi}{k^2}[kz_0-\sin(kz_0)]\\
&\times 2\frac{\cos\left(\pi - \frac{z_0 x}{a}\right)+f_{1}(kz_0)}{\cos(kz_0)-\cos\left(\pi - \frac{z_0 x}{a} \right)}\\
&\simeq-\frac{i\xi}{k^2}\frac{1}{3}(kz_0)^3 \frac{2-\cos\left(\frac{z_0x}{a}\right)}{1+\cos\left(\frac{z_0x}{a}\right)},
\end{aligned}
\end{equation}
where we have used the fact that $\lim_{y\to 0}f_{1}(y)=2$ [with $f_{1}(y)$ defined in eqn. (33) of \cite{anastasiscnt}] and the small-argument expansion $\sin{x}\sim x -(1/3!)x^3$.

We also recast \ref{eq:Bdef} in a different form, which will prove helpful for the asymptotic study, as below:
\begin{equation}\label{eq:Bexpr}
B\left(\pi - \frac{z_0 x}{a}\right)=-i \frac{V}{\zeta_0}k z_0^2\frac{\rho}{z_0}\sum_{m=-\infty}^{\infty}\frac{1}{r_{m}}K_{1}\left(\frac{\rho}{z_0}r_{m}\right),
\end{equation}
with $r_{m} \simeq \left|(2m-1)\pi -(z_0 x)/a\right|$ for $k z_0 \ll 1$. 

Having found the main quantities appearing in \ref{eq:Icorr2}, we will now focus our attention on two main cases, in the spirit of \cite{EffCurrent2013}:

\subsubsection{Radial distance of the order of the segmentation length, $\rho/z_0=\mathcal{O}(1)$ }

Replacing the argument $z_0 x/a$ by zero in all instances, the integrand in \ref{eq:Icorr2} assumes the form
\begin{equation}\label{eq:fintegrand}
f(\rho, z_0;x) \simeq \frac{iV}{\zeta_0}\frac{\pi^5}{16}\eta\left(\frac{\rho}{z_0}\right)\frac{1}{\cosh^2(x)}\frac{i\xi}{k}\frac{(kz_0)^3}{6} \cos\left(\frac{nz_0x}{a}\right),
\end{equation}
where the function $\eta(y)$ is defined as
\begin{equation}\label{eq:Etadef}
\eta(y)\equiv \frac{4y}{\pi}\sum_{m=-\infty}^{\infty}\frac{1}{\left|2m-1\right|}K_{1}\left(\left|2m-1 \right|\pi y \right),
\end{equation}
and is discussed extensively in \cite{EffCurrent2013}. Then, replacing $a/z_0$ by infinity in the upper limit of the second integral of \ref{eq:Icorr2}, we obtain
\begin{equation}\label{eq:effwithI}
\begin{aligned}
I_{\rm eff,\,corr}^{(\infty)}(\rho, nz_0,z_0)&\simeq\frac{(-1)^{n}}{2}\,e^{2\pi a/z_0} \frac{V}{\zeta_0}\frac{\pi^5}{16}\\
&\times\eta\left(\frac{\rho}{z_0}\right)\frac{\xi}{k}\frac{(kz_0)^3}{6} \int_{0}^{\infty}\frac{\cos\left(\frac{nz_0 x}{a}\right)}{\cosh^2(x)}\,dx.
\end{aligned}
\end{equation}
This correction is manifestly a real quantity. For the calculation of the integral in \ref{eq:effwithI} we use Entry 2.5.48.5 of \cite{Prudnikov} to write
\begin{equation*}
\int_{0}^{\infty}\frac{\cos\left(\frac{nz_0 x}{a}\right)}{\cosh^2(x)}\,dx=\frac{u}{\sinh{u}},
\end{equation*}
with $u=n\pi z_0/(2a)$; hence for $\rho=0$ the effective-current correction is (with $\eta(0)=1$)
\begin{equation}\label{eq:FinalExpr}
I_{\rm eff,\,corr}^{(\infty)}(0, nz_0,z_0) \simeq (-1)^{n}\,e^{2\pi a/z_0} \frac{V}{\zeta_0}\frac{\pi^5}{16}\frac{\xi}{k}\frac{(kz_0)^3}{12} \frac{u}{\sinh{u}},
\end{equation}
This expression is identical to that of eqn. (48) in \cite{anastasiscnt} for $z_0/a=0$ [apart from a factor of two owing to the different numerical method employed here (see Sec. VIII of \cite{Hallen2001})], verifying that $I_{\rm eff}^{(\infty)}(0, nz_0,z_0)$ equals the current $I_{n}^{(\infty)}$ obtained by the numerical method with the approximate kernel \cite{EffCurrent2011, EffCurrent2013, anastasiscnt}. The oscillations around the driving point $z=0$ predicted by the factor $(-1)^{n+1}$ decrease rapidly in magnitude with increasing $z_0/a$. This behavior is shown for the current ${\rm Re}\{I_{\rm full, ap;\,n}/V\}$ for the finite antenna in Figs. \ref{fig:Figure1}(a,c). The prediction of \ref{eq:FinalExpr} (for $\rho=0$) depicted in Fig. \ref{fig:Figure1}(d) is in good agreement with ${\rm Re}\{I_{\rm full, ap;\,n}/V\}$ for a finite-length antenna, as shown in Fig. \ref{fig:Figure1}(c) depicting the central oscillations owing to finite conductivity.

From the expression for the total effective asymptotic current at a nonzero $\rho \sim z_0$, up to first order in $\xi/k$ [see eqn. (2.17) of \cite{EffCurrent2013} for the zero-order term corresponding to the PC antenna]
\begin{equation}\label{eq:IeffRho}
\begin{aligned}
& I_{\rm eff}^{(\infty)}(\rho, nz_0,z_0) \simeq (-1)^{n+1}  \frac{\pi^3}{16}\frac{V}{\zeta_0}(kz_0)\,\exp\left(\pi \frac{a}{z_0}\right) \eta\left(\frac{\rho}{z_0} \right)\\
& \times \Bigg[\frac{i}{\sqrt{2}}\,  \sqrt{\frac{z_0}{a}}\frac{1}{\cosh{u}}
- \pi^2\frac{\xi}{k}\frac{(kz_0)^2}{12}\exp\left(\pi \frac{a}{z_0}\right) \frac{u}{\sinh{u}} \Bigg],
\end{aligned}
\end{equation}
we conclude that \textit{the first-order correction has the same form as that of eqn. (48) in \cite{anastasiscnt} for $z_0/a=0$, multiplied by the function $\eta(\rho/z_0)$ to include the dependence of the effective current on the distance $\rho$ from the $z$-axis.}

\subsubsection{Radial distance much larger than the segmentation length, with $z_0 \ll \rho \ll a$} 

We retain the two dominant terms in the series of the expression \ref{eq:Bexpr} for $B$, namely those with $m=0$ and $m=1$, as in [eqn. (2.22) of \cite{EffCurrent2013}]:
\begin{equation}\label{eq:dominant}
\begin{aligned}
&B\left(\pi - \frac{z_0 x}{a}\right)\simeq -i \frac{V}{\zeta_0}k z_0^2 \frac{\rho}{z_0}\\
&\times\left[\frac{K_{1}\left(\frac{\rho}{z_0}\left(\pi + \frac{z_0 x}{a}\right)\right)}{\pi+z_0x/a} + \frac{K_{1}\left(\frac{\rho}{z_0}\left(\pi - \frac{z_0 x}{a}\right)\right)}{\pi-z_0x/a} \right].
\end{aligned}
\end{equation}
We can further employ the large-argument approximation of the modified Bessel function $K_{1}(y) \sim \sqrt{\pi/(2y)}\,e^{-y}$, to write
\begin{equation*}
\begin{aligned}
&B\left(\pi - \frac{z_0 x}{a}\right)\simeq -i \frac{V}{\zeta_0} \sqrt{\frac{\pi}{2}} k z_0^2 \sqrt{\frac{\rho}{z_0}}e^{-\pi \frac{\rho}{z_0}}\\
&\times \left[\frac{e^{-\rho x/a}}{(\pi+z_0x/a)^{3/2}} + \frac{e^{\rho x/a}}{(\pi-z_0x/a)^{3/2}} \right].
\end{aligned}
\end{equation*}
We now set $z_0/a=0$ in the expression \ref{eq:Icorr2} for $I^{(\infty)}_{\rm eff, \, corr}(\rho, nz_0,z_0)$, while keeping the remaining arguments $ka, nz_0/a$ and $\rho/z_0$ fixed. Integrating from $0$ to $\infty$ yields
\begin{equation}\label{eq:Case2withJ2}
\begin{aligned}
I_{\rm eff;\,n}^{(\infty)}(\rho, nz_0,z_0)& \simeq \frac{(-1)^{n}}{2\sqrt{2}}\frac{V}{\zeta_0}\pi^4\sqrt{\frac{\rho}{z_0}} \frac{\xi}{k} \frac{1}{12}(kz_{0})^3 e^{\frac{\pi}{z_0}(2a-\rho)}\\
& \times \int_{0}^{\infty}\frac{\cosh\left(\frac{\rho x}{a}\right)}{\cosh^2 x}\cos\left(\frac{nz_0 x}{a}\right)\, dx.
\end{aligned}
\end{equation}
Using Entry 2.5.48.12 of \cite{Prudnikov} for the calculation of the integral $\int_{0}^{\infty}\frac{\cosh\left(\frac{\rho x}{a}\right)}{\cosh^2 x}\cos\left(\frac{nz_0 x}{a}\right)\, dx$ in \ref{eq:Case2withJ2} yields the final result for the total effective asymptotic current at $\rho \gg z_0$, up to first order in $\xi/k$ [see eqn. (2.26) of \cite{EffCurrent2013} for the zero-order term corresponding to the PC antenna]
\begin{equation}\label{eq:Case2form}
\begin{aligned}
&I_{\rm eff;\,n}^{(\infty)}(\rho, nz_0,z_0) \simeq (-1)^{n+1}\frac{\pi^2 V}{2\zeta_0}\sqrt{\frac{\rho}{z_0}}(kz_0)\,e^{\frac{\pi}{z_0}(a-\rho)} \\
&\times \Bigg[i\sqrt{\frac{z_0}{a}}\frac{\cos{v} \cosh{u}}{\cosh(2u) + \cos(2v)}-\frac{\pi^2e^{\frac{\pi a}{z_0}}}{6\sqrt{2}}\frac{\xi}{k}(kz_{0})^2\\
&\times\frac{u\cos{v}\sinh{u}+v\sin{v}\cosh{u}}{\cosh(2u)-\cos(2v)}\Bigg],
\end{aligned}
\end{equation} 
with $u=n\pi z_0/(2a)$ and $v=\pi \rho/(2a)$. From this expression we conclude that \textit{at larger radial distances the effective current continues to oscillate along the $z$-axis, however it has greatly reduced in magnitude from its value near $\rho=0$. This behavior is reminiscent of a surface wave, as explained in detail (for the perfectly conducting case) in Sec. 4 of \cite{EffCurrent2013}.} 

\begin{figure}[!h]
\centering
\includegraphics[width=0.98\columnwidth]{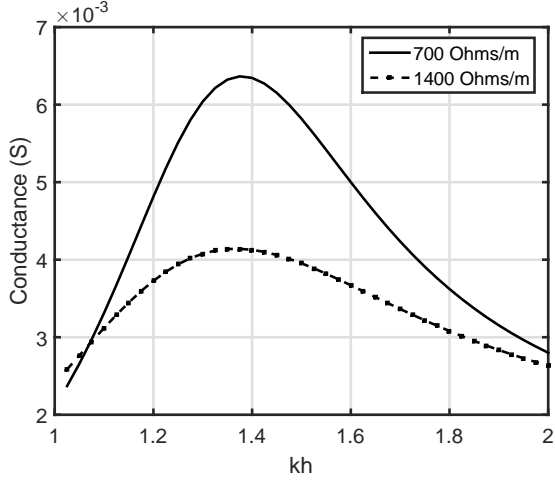}
\caption{Effective input conductance $G$ of a resistive cylindrical dipole with varying $kh$, after Fig. 1 of \cite{PopovicTAP}, calculated from \ref{eq:IeffInFin} at $\rho=a$ as ${\rm Re}\{I_{\rm eff;\,0}/V\}$ at the driving point $z=0$ for internal impedance per unit length $z_{i}=700\,$Ohms$/m$ (solid line) and $z_{i}=1400\,$Ohms$/m$ (broken line). The parameter $\xi$ is related to $z_{i}$ via $\xi=z_{i}/(2\zeta_0)$. Other parameters: $h/a=35.7608$, $a=0.32\,$cm and $N=80$.}	
\label{fig:Figure3}
\end{figure}

\subsection{Limiting case of vanishing conductance}

We will now demonstrate that the effective current has a meaningful small-$z_0$ limit of physical significance for $0<\rho\leq a$, in the sole presence of the loss kernel [following the steps of Sec. 4 in \cite{EffCurrent2011}]. For that we need the FT of the loss kernel [see eqn. (11) of \cite{anastasiscnt}]
\begin{equation}\label{eq:KlossFourier}
\overline{K}_{\rm loss}(\zeta)=\frac{2ik\xi}{k^2-\zeta^2}.
\end{equation}
Following the series of approximations leading to eqn. (4.4) in \cite{EffCurrent2011} for $\overline{K}=\overline{K}_{\rm loss}(\zeta)$, we end up in the following expression for $z_0 \to 0$
\begin{equation}\label{eq:IHertz}
\begin{aligned}
&I_{\rm eff, loss}^{(\infty)}(\rho,z,z_0)\sim \frac{V}{2\pi\zeta_{0}}\int_{0}^{\infty}\cos(\zeta z)\frac{\overline{L}(\zeta,\rho)}{\overline{K}_{\rm loss}(\zeta)}\, d\zeta \\
&=\frac{V}{2\pi\zeta_{0}}\int_{0}^{\infty}\cos(\zeta z)\frac{\rho}{\xi}\sqrt{\zeta^2-k^2}\,K_{1}\left(\rho\sqrt{\zeta^2-k^2}\right)\, d\zeta\\
&=2\pi \rho \frac{I_0 z_0}{4\pi}(ik)\left(\frac{1}{ikr}-1\right)\frac{e^{ikr}}{r}\sin\theta,
\end{aligned}
\end{equation}
with $r=\sqrt{\rho^2+z^2}$, $\sin\theta=\rho/r$ and $I_0 z_0=V/(2\zeta_0 \xi)$, the coefficient of a $\delta(z)$ current distribution solving the integral equation, corresponding to eqn. \ref{eq:Pocklington}, for the infinite antenna with $K=K_{\rm loss}$. In the derivation of \ref{eq:IHertz} we have used eqn. (3.23) of \cite{EffCurrent2011} and eqn. (29) of \cite{anastasiscnt} (see Appendix B for more details). \textit{The resulting expression is $2\pi \rho$ times the magnetic field of a Hertzian dipole of infinitesimal length $z_0$, in accordance with the definition of the effective current.}

For large values of $\xi/k$, when the perturbative approach described in Subsec. \ref{subsec:largecond} breaks down, the effective current in the sole presence of the loss kernel is instead given by the formula [see eqn. (17) of \cite{anastasiscnt}]
\begin{equation}\label{eq:IcorrAloss}
\begin{aligned}
I_{\rm eff, loss}^{(\infty)}(\rho, nz_0,z_0)&=\frac{z_0}{2\pi}\int_{-\pi/z_0}^{\pi/z_0}\frac{B(z_0 \zeta, \rho)}{A_{\rm loss}(z_0 \zeta)}e^{-i n z_0 \zeta}\, d\zeta\\
&=\frac{1}{2\pi}\int_{-\pi}^{\pi}\frac{B(\theta, \rho)}{A_{\rm loss}(\theta)}\cos(n\theta)d\theta,
\end{aligned}
\end{equation}
For $\rho \to 0$ we use the small-argument approximation $K_{n}(y) \sim \frac{(n-1)!}{2}\, (y/2)^{-n}$ (for $y \ll n$), to write
\begin{equation*}
\begin{aligned}
B(z_0 \zeta, \rho)&\simeq -i \frac{V}{\zeta_0}\frac{k z_0^2}{4\pi^2}\sum_{m=-\infty}^{\infty}\frac{1}{[m-\zeta z_0/(2\pi)]^2}\\
&=-i \frac{V}{\zeta_0}\frac{k z_0^2}{4}\frac{1}{\sin^2 (\zeta z_0/2)}.
\end{aligned}
\end{equation*}
For small $kz_0 \to 0$ we can write
\begin{equation*}
\frac{B(z_0 \zeta, \rho)}{A_{\rm loss}(\zeta z_0)}\simeq\frac{3}{2 z_0}\frac{V}{\zeta_0 \xi}\frac{1}{2+\cos(\zeta z_0)},
\end{equation*}
such that [setting $\theta=\zeta z_0$ and invoking eqn. \ref{eq:IcorrAloss}]
\begin{equation}\label{eq:IeffDelta1}
\begin{aligned}
I_{\rm eff, loss}^{(\infty)}(0, nz_0,z_0)&\simeq \frac{3}{4\pi z_0}\frac{V}{\zeta_0 \xi} \int_{-\pi}^{\pi}\frac{\cos(n\theta)}{2+\cos{\theta}}d\theta\\
&=\frac{\sqrt{3}}{2}\frac{V}{\zeta_0 \xi z_0}(-1)^{n}(2+\sqrt{3})^{-|n|}.
\end{aligned}
\end{equation}
We now consider the sum
\begin{equation}\label{eq:deltaseq}
\begin{aligned}
\sum_{n=-\infty}^{\infty}z_0 I_{\rm eff, loss}^{(\infty)}(0, nz_0,z_0)&=\frac{3V}{4\pi\zeta_0 \xi} \int_{-\pi}^{\pi}\frac{\sum_{n=-\infty}^{\infty}\cos(n\theta)}{2+\cos{\theta}}d\theta\\
&=\frac{3}{2}\frac{V}{\zeta_0 \xi} \frac{1}{3}=\frac{V}{2\zeta_0 \xi},
\end{aligned}
\end{equation}
where we have made use of the identity
\begin{equation*}
\sum_{n=-\infty}^{\infty} \cos(n \theta)=2\pi \sum_{m=-\infty}^{\infty} \delta(\theta-2m \pi).
\end{equation*}
We have then proved that the effective current converges to a delta-function solution for $kz_0 \ll 1$ and $\rho \to 0$, as required for the small-$z_0$ limit distribution [see eqn. (40) of \cite{anastasiscnt}]. Note that the delta-sequence derived here is essentially different from that of eqn. (41) of \cite{anastasiscnt}.

\section{Discussion of results and relevance to experiments}

The asymptotic results derived in Sec. \ref{sec:effectcur} for the imperfectly conducting infinite antenna can serve as a useful guide for the numerical investigation of unphysical current oscillations and their alleviation for a finite-length antenna when $N \gg h/a$. In Fig. \ref{fig:Figure1} we depict an intensely oscillating distribution ${\rm Re}\{I_{\rm full, ap;\,n}/V\}$ resulting from the application of MoM-A for a small antenna with $h/\lambda=1/20$ and  $\xi \lambda \sim 10^{-9}$. Here the oscillating and the smoothed distributions for the real part differ by eight orders of magnitude. This figure draws inspiration from Table III of \cite{anastasiscnt}, bearing direct relevance to the current distribution on CNT antennas. The smoothed (effective) current (drawn in $\circ$) is relatively close to the exact-current distribution, an agreement which betters significantly with increasing antenna length [as evidenced in Figs. \ref{fig:Figure2} (c,d)].  In Fig. \ref{fig:Figure1}(a) the oscillations around $z=0$ in the real part dominate over the oscillations at the two ends for $z_0/a=0.125$. Increasing the ratio $h/\lambda$ from $1/20$ to $1/15$ has the effect of reducing dramatically the central oscillations, which now become comparable to the side oscillations for $z_0/a=1/6$ [compare Figs. \ref{fig:Figure1} (a,c)] due to the presence of the approximate kernel $K_{\rm ap}$. The reduction of the central oscillations is also a prediction of the asymptotic formula \ref{eq:FinalExpr}. The drive susceptance predicted from the smoothed current distribution, which does not depend on $N$ for $h/\lambda$ fixed, shows a disparity with respect to the exact-kernel result, a difference diminishing with increasing $h/\lambda$ for fixed $N$ and $a$.

For the CNTs discussed in \cite{FundamentalTransm,keller2014electromagnetic, huang2008performance, fichtner2008investigation} the unphysical oscillations arising from the non-solvability of Hall\'{e}n's equation with the approximate kernel for a perfectly conducting antenna occur for very high values of $N \gg 10^3$, owing to the extremely small radius to wavelength ratio of a typical CNT. At the same time, since $a/z_0\ll 1$, the central oscillations in the real part of the current distribution due to the finite conductivity are negligible for typical values of $\xi \lambda$ in CNTs, as eqn. \ref{eq:FinalExpr} demonstrates. On the other hand, a particular example of a realistic CNT with large $a/\lambda$ \cite{xiapeiro} is discussed in detail in \cite{anastasiscnt} (see Tables I and II), where oscillations do in fact occur for $N=125$. This is also the case for a finite-conductivity antenna with $h/a \sim 10$ and $N=80$, as depicted in Fig. \ref{fig:Figure1} of our work.

The effective current formulation pertains as well to the case of a resistive dipole, as discussed in [Table I and Fig. 2 of \cite{PopovicTAP}], where oscillations for the full approximate-kernel distribution around $z=0$ differ by one order of magnitude for the real and imaginary parts, as shown in Figs. \ref{fig:Figure2} (a, b) for a half-wavelength antenna. Both distributions are smoothed efficiently, as depicted in Figs. \ref{fig:Figure2} (c,d), yielding a very good agreement with the exact-kernel distribution $({\rm Re, Im})\{I_{\rm full, ex;\,n}/V\}$. In comparison to the dipole of Fig. \ref{fig:Figure1}, here a much larger value of $\xi \lambda \sim 10^{-1}$ has been used; however, the central oscillations for ${\rm Re}\{I_{\rm full, ap;\,n}/V\}$ [shown in Fig. \ref{fig:Figure2}(a)] are much weaker, since $z_0/a\simeq 0.4$. 

Results for the input admittance $G={\rm Re}(Y)$ of resistive dipoles with a fixed internal impedance per unit length, $z_{i}$, compare very well to the experimental data obtained for varying wavelength of operation (and $\xi \lambda \sim 10^{-1}$). This is evident from a comparison between Fig. \ref{fig:Figure3} in our work, and Fig. 1 of \cite{PopovicTAP}. In the former we depict ${\rm Re}\{I_{\rm eff}(n=0)/V\}$ calculated from Method MoM-A and the application of \ref{eq:IeffInFin} for $\rho=a$, following the smoothing of an oscillating distribution around the driving point $z=0$ for the real part of $I_{\rm full, ap;\,n}/V$ . The attained alleviation of oscillations is qualitatively similar to the one depicted in Figs. \ref{fig:Figure2}(a, c), in very good agreement with ${\rm Re}\{I_{\rm full, ex;\,n}/V\}$. 

\section{Conclusions}

We have presented an {\it effective-current method} for post-processing the oscillating solutions of Hall\'{e}n's equation with the approximate kernel for an imperfectly conducting linear cylindrical antenna, in order to obtain a smooth current that is close to the one obtained with the exact kernel. The agreement betters with increasing the antenna length for a fixed $N$, as a result of a larger ratio $z_{0}/a$. The presence of finite conductance entails additional oscillations for the real part as a perturbative correction to the solution for the PC antenna. Their strength, however, can be appreciable, depending on $z_0/a$. In the limit of zero conductance, the resulting distribution is a rapidly-oscillating delta sequence; this behavior is reproduced by the effective current close to the axis of symmetry.

Future work could enlarge the scope of the effective-current method to account for other numerical methods in conjunction with analytical studies for the infinite antenna. Solutions to Hall\'{e}n's equation where the exact kernel is used could also be sequestered, taking into account the well-known logarithmic divergence for the perfectly conducting antenna, which is captured as well by the effective current \cite{EffCurrent2011}. 

\begin{center}
*****
\end{center}

\newpage

 \bibliography{Paper_0618_ThA_ArXiv}

\onecolumngrid
\appendix

\section{Derivation of the delta-sequence in eqn. \ref{eq:IeffDelta1}}

The integral featuring in eqn. \ref{eq:IeffDelta1} is evaluated using the residue theorem following the substitution $z=e^{i\theta}$, as below:
\begin{equation}
\begin{aligned}
C_{n}&\equiv\int_{-\pi}^{\pi}\frac{\cos(n\theta)}{2+\cos{\theta}}d\theta=\frac{1}{i}\oint_{|z|=1}\frac{2z^{|n|}}{z^2+4z+1}dz\\
&=\frac{2\pi i}{i} 2 \frac{r_{+}^{|n|}}{r_{+}-r_{-}},
\end{aligned}
\end{equation}
where $r_{\pm}=-2\pm \sqrt{3}$. The final result is
\begin{equation*}
C_{n}=2\frac{2\pi (-1)^{n}(2-\sqrt{3})^{|n|}}{2\sqrt{3}}=\frac{2\pi}{\sqrt{3}}(-1)^{n}(2+\sqrt{3})^{-|n|},
\end{equation*}
whence
\begin{equation}
\begin{aligned}
I_{\rm eff, loss}^{(\infty)}(0, nz_0,z_0)&\simeq \frac{3}{4\pi z_0}\frac{V}{\zeta_0 \xi}\frac{2\pi}{\sqrt{3}}(-1)^{n}(2+\sqrt{3})^{-|n|}\\
&=\frac{\sqrt{3}}{2}\frac{V}{\zeta_0 \xi}(-1)^{n}(2+\sqrt{3})^{-|n|}.
\end{aligned}
\end{equation}

\section{Derivation of the Hertzian dipole field of eqn. \ref{eq:IHertz}}\label{ap:BHertz}

We use the FT of $\exp(ik\sqrt{z^2+\rho^2})$ (with $\mathrm{Im}(k)>0$)
\begin{equation*}
\begin{aligned}
\overline{L}(\zeta, \rho)&=\int_{-\infty}^{\infty}\exp(ik\sqrt{z^2+\rho^2})\exp(i\zeta z)\, dz\\
&=\frac{2k\rho}{i\sqrt{\zeta^2-k^2}}\, K_{1}\left(\rho\sqrt{\zeta^2-k^2}\right),
\end{aligned}
\end{equation*}
featuring in the definition \ref{eq:Bdef}, to write (for $z_0 \to 0$)
\begin{equation}
\begin{aligned}
&I_{\rm eff, loss}(\rho,z,z_0)\sim \frac{V}{2\pi\zeta_{0}}\int_{0}^{\infty}\cos(\zeta z)\frac{\overline{L}(\zeta,\rho)}{\overline{K}_{\rm loss}(\zeta)}\, d\zeta\\
&=\frac{V}{2\pi\zeta_{0}}\int_{0}^{\infty}\cos(\zeta z)\frac{\rho}{\xi}\sqrt{\zeta^2-k^2}\,K_{1}\left(\rho\sqrt{\zeta^2-k^2}\right)\, d\zeta\\
&=\frac{V}{4\pi\zeta_{0}}\frac{i}{2k\rho}\int_{-\infty}^{\infty}\frac{\rho}{\xi}\frac{2k\rho}{i}\frac{\zeta^2-k^2}{\sqrt{\zeta^2-k^2}}\,K_{1}\left(\rho\sqrt{\zeta^2-k^2}\right)e^{-i\zeta z} d\zeta,
\end{aligned}
\end{equation}
giving
\begin{equation}
\begin{aligned}
&I_{\rm eff, loss}(\rho,z,z_0)\sim\frac{V}{4i k \zeta_0}\frac{1}{\xi}\left(\frac{\partial^2}{\partial z^2} +k^2 \right)\frac{1}{2\pi}\int_{-\infty}^{\infty}\overline{L}(\zeta, \rho)e^{-i\zeta z} d\zeta\\
&=\frac{V}{4i \pi k \zeta_0}\frac{1}{\xi}\left(\frac{\partial^2}{\partial z^2} +k^2 \right)\exp(ik\sqrt{z^2+\rho^2})\\
&=\frac{V}{4i k \zeta_0}\frac{1}{\xi}\exp(ik\sqrt{z^2+\rho^2})\left[k^2\frac{{\rho}^2}{z^2+\rho^2}+ik \frac{\rho^2}{(z^2+\rho^2)^{3/2}}\right]\\
&=\frac{V}{4i k \zeta_0}\frac{\rho}{\rho\xi}\sin^2 \theta\left(k^2  + ik \frac{1}{r}\right)e^{ikr}\\
&=2\pi \rho \frac{V}{8i \pi k \zeta_0 \xi}\sin \theta\left(k^2 \frac{1}{r} + ik \frac{1}{r^2}\right)e^{ikr}.
\end{aligned}
\end{equation}
With the identification $I_0=V/(2\zeta_0 \xi z_0)$, the expression \ref{eq:IHertz} readily follows.

\end{document}